\documentclass[12pt]{article}

\def\beq{\begin{eqnarray}}
\def\eeq{\end{eqnarray}}

\def\L*{{\cal L}_*}
\def\lsim{\mathrel{\rlap{\lower3pt\hbox{\hskip0pt$\sim$}}
     \raise1pt\hbox{$<$}}}         %less than or approx. symbol
\def\gsim{\mathrel{\rlap{\lower4pt\hbox{\hskip1pt$\sim$}}
     \raise1pt\hbox{$>$}}}         %greater than or approx. symbol

\usepackage{amsmath}
\usepackage{amsfonts}

\begin{document}
\begin{titlepage}

\centerline{\Large \bf Massless Limit of Gravitational Higgs Mechanism}
\medskip

\centerline{\large Zurab Kakushadze}

\bigskip

\centerline{\em 200 Rector Place, Apt 41F, New York, NY 10280}
\centerline{\tt zura@kakushadze.com}
\centerline{(November 11, 2007)}

\bigskip
\medskip

\begin{abstract}
{}In gravitational Higgs mechanism graviton components acquire mass via spontaneous diffeomorphism breaking
by scalar vacuum expectation values. We point out that in the massless limit the resulting theory is not
Einstein-Hilbert gravity (EHG) but constrained gravity (CG). Consequently, massive solutions in the massless limit must be compared to those
in CG (as opposed to EHG). We discuss spherically symmetric solutions in this context.
The Schwarzschild solution in EHG can be coordinate-transformed such that it is also a solution in CG.
The resulting solutions are non-perturbative in the asymptotic regime, and are reproduced in the massless limit of asymptotic massive solutions, hence
no van Dam-Veltman-Zakharov discontinuity. We point out that higher curvature terms must be included to obtain non-singular
spherically symmetric massive solutions and discuss a suitable framework.
\end{abstract}
\end{titlepage}

\newpage

\section{Introduction and Summary}

{}In gravitational Higgs mechanism \cite{KL} graviton components acquire mass via spontaneous breaking of diffeomorphisms
by scalar vacuum expectation values\footnote{For earlier and subsequent related works, see, {\em e.g.}, \cite{Duff, OP, GT, Siegel, Por, AGS, Ch, Ban, Lec, Kir, Kiritsis, Tin}.
For a recent review of massive gravity in the context of infinite volume extra dimensions, see
\cite{Gab} and references therein. For a recent review of spontaneous breaking of diffeomorphism symmetry
in the context of Lorentz violating Chern-Simons modification of gravity, see \cite{Jackiw} and references
therein.}. Recently, gravitational Higgs mechanism was revisited in the context of obtaining massive
gravity directly in four dimensions \cite{thooft, ZK, Oda, 0710.1061}. A general Lorentz invariant mass term for the graviton $h_{MN}$ is of the form
\begin{equation}
 -{m^2\over 4} \left[h_{MN}h^{MN} - \beta (h^M_M)^2\right]~,
\end{equation}
where $\beta$ is a dimensionless parameter. For $\beta\neq 1$ the trace component $h^M_M$,
which is a ghost, is a propagating degree of
freedom, while it decouples for the Pauli-Fierz mass term with $\beta = 1$. Massive gravity in Minkowski space with $\beta = 1$ can be obtained
via gravitational Higgs mechanism by including higher derivative terms in the scalar sector and appropriately tuning the cosmological constant
\cite{0710.1061}.

{}The framework of \cite{0710.1061} provides a ghost-free, non-linear and fully covariant definition of massive gravity
via gravitational Higgs mechanism with spontaneously (as opposed to explicitly) broken diffeomorphisms.
In this paper we discuss the massless limit of gravitational Higgs mechanism within this framework.
Our key observation is that in the massless limit the resulting theory is not
Einstein-Hilbert gravity (EHG) but constrained gravity (CG). Consequently, massive solutions in the massless limit must be compared to those
in CG (as opposed to EHG). For example, the usual Schwarzschild solution in EHG does not satisfy the constraint in CG, so we cannot compare
massive solutions in the massless limit to the Schwarzschild solution in EHG. On the other hand,
the Schwarzschild solution in EHG can be coordinate-transformed such that it is also a solution in CG.
The resulting solutions are non-perturbative in the asymptotic regime, and are reproduced in the massless limit of asymptotic massive solutions. Therefore,
there is no van Dam-Veltman-Zakharov (vDVZ) discontinuity \cite{vDV, Zak} in the massless limit of gravitational Higgs mechanism in the spirit of \cite{Vain}\footnote{In the context of brane induced gravity no vDVZ discontinuity was argued in \cite{DDGV, Gruz, DGZ, GI} by studying non-perturbative solutions.}.

{}We also study massive solutions in the limit corresponding to the near-horizon regime in massless solutions. We point out that in the lowest-curvature
approximation (Einstein-Hilbert term) massive solutions have a naked singularity. This implies that the lowest-curvature approximation breaks down,
and higher curvature terms must be included to obtain non-singular solutions. Adding higher curvature terms does not modify the no vDVZ conclusion in the asymptotic
regime. We discuss a framework for constructing non-singular solutions for all values of the graviton mass in the context of gravitational Higgs mechanism.

\section{Gravitational Higgs Mechanism}

{}In this section we review the construction of \cite{0710.1061}, and discuss the massless limit of gravitational Higgs mechanism.
We consider gravity in $D$ dimensions coupled to scalar fields $\phi^A$,
$A = 0,\dots, D-1$. We will take the metric $Z_{AB}$ for the global scalar indices $A,B,\dots$ to be the Minkowski metric:
$Z_{AB} = \eta_{AB}$.

{}Coordinate-dependent scalar VEVs break diffeomorphisms spontaneously. This results in massive gravity. To obtain a unitary
theory ({\em i.e.}, the Pauli-Fierz mass term for the graviton in the linearized theory), we must include higher derivative terms in
the scalar sector. In fact, adding four-derivative terms suffices, but for now we will keep our discussion here more general.

{}Thus, consider the following action:
\begin{equation}
 S_Y = M_P^{D-2}\int d^Dx \sqrt{-G}\left[ R - V(Y)\right]~,
 \label{actionphiY}
\end{equation}
where {\em a priori} $V(Y)$ is a generic function of $Y$, and
\begin{eqnarray}\label{Y}
 &&Y_{MN} \equiv Z_{AB} \nabla_M\phi^A \nabla_N\phi^B~,\\
 &&Y\equiv Y_{MN}G^{MN}~.
\end{eqnarray}
{}The equations of motion read:
\begin{eqnarray}
 \label{phiY}
 && \nabla^M\left(V^\prime(Y)\nabla_M \phi^A\right) = 0~,\\
 \label{einsteinY}
 && R_{MN} - {1\over 2}G_{MN} R = V^\prime(Y) Y_{MN}
 -{1\over 2}G_{MN} V(Y)~,
\end{eqnarray}
where prime denotes derivative w.r.t. Y. Multiplying (\ref{phiY}) by $Z_{AB} \nabla_S\phi^B$ and contracting indices, we
can rewrite the scalar equations of motion as follows:
\begin{equation}\label{phiY.1}
 \partial_M\left[\sqrt{-G} V^\prime(Y) G^{MN}Y_{NS}\right] - {1\over 2}\sqrt{-G} V^\prime(Y) G^{MN}\partial_S Y_{MN} = 0~.
\end{equation}
Since the theory possesses full diffeomorphism symmetry, (\ref{phiY.1}) and (\ref{einsteinY}) are not all independent but linearly related
due to Bianchi identities. Thus, multiplying (\ref{einsteinY}) by $\sqrt{-G}$, differentiating w.r.t. $\nabla^N$ and contracting indices we arrive
at (\ref{phiY.1}).

{}We are interested in finding solutions of the form:
\begin{eqnarray}\label{solphiY}
 &&\phi^A = m~{\delta^A}_M~x^M~,\\
 \label{solGY}
 &&G_{MN} = \eta_{MN}~.
\end{eqnarray}
These exist for a class of ``potentials" $V(Y)$ such that the following equation
\begin{equation}\label{tune-V.0}
 Y~V^\prime(Y) = {D\over 2} ~V(Y)
\end{equation}
has non-trivial solutions\footnote{This condition is not particularly constraining. Such solutions exist even for linear potentials $V = \Lambda + Y$.
In this case the restriction is that the cosmological constant be negative.}. Let us denote such a solution by $Y_*$. Then we have
\begin{equation}
 m^2 = Y_* / D~.
\end{equation}
Note that (\ref{tune-V.0}) is not a tuning condition on couplings in $V$. Rather, it fixes the scalar VEVs
(whose slopes are given by $m$) in terms of various couplings in $V$.

{}The equations of motion (\ref{phiY}) and (\ref{einsteinY}) are invariant under the full diffeomorphisms of the theory
($G_{MN}\equiv \eta_{MN} + h_{MN}$):
\begin{eqnarray}\label{diffphiY}
 &&\delta\varphi^A =\nabla_M \phi^A \xi^M = m~{\delta^A}_M ~\xi^M~,\\
 &&\delta h_{MN} = \nabla_M \xi_N + \nabla_N \xi_M~.
\end{eqnarray}
The scalar fluctuations $\varphi^A$ around the background (\ref{solphiY}) can be gauged away (set to zero)
using the diffeomorphisms (\ref{diffphiY}). Once we gauge away the scalars, diffeomorphisms
can no longer be used to gauge away any of the graviton components $h_{MN}$. After setting $\varphi^A = 0$,
in the linearized approximation we have ($h\equiv h_M^M$):
\begin{eqnarray}
 &&\partial^N h_{MN} - \beta \partial_M h  = 0~,\\
 \label{linEOM1}
 &&R_{MN} - {1\over 2}G_{MN} R = {M_h^2\over 2} \left(\beta\eta_{MN}h  - h_{MN} \right)~,
\end{eqnarray}
where
\begin{eqnarray}
 &&\beta \equiv {1\over 2} - 2m^4 {V^{\prime\prime}(Y_*)\over{V(Y_*)}}~,\\
 &&M_h^2\equiv V(Y_*)~.
\end{eqnarray}
In particular, for a special class of potentials with
\begin{equation}\label{tune-V}
 4Y_*^2 V^{\prime\prime}(Y_*) = -D^2 V(Y_*)~,
\end{equation}
we have $\beta = 1$, and the correct tensor structure for massive gravity without non-unitary states. In this case
the explicit form of the equations of motion is given by:
\begin{eqnarray}
 &&h = 0~,\\
 &&\partial^N h_{MN} = 0~,\\
 &&\partial^S\partial_S h_{MN} = M_h^2 h_{MN}~,
\end{eqnarray}
In particular, the ghost state $h$ decouples, and we have massive gravity with $(D+1)(D-2)/2$ propagating degrees of freedom.

{}Thus, we can get the Pauli-Fierz combination of the mass terms
for the graviton if we tune {\em one} combination of couplings. In fact, this tuning is nothing but tuning of
the cosmological constant - indeed, (\ref{tune-V}) relates the cosmological constant to higher derivative couplings.

{}Thus, consider a simple example:
\begin{equation}\label{quartic}
 V = \Lambda + Y + \lambda Y^2~.
\end{equation}
The first term is the cosmological constant, the second term is the kinetic term for the scalars (which can always be normalized
such that the corresponding coefficient is 1), and the third term is a four-derivative term. We then have:
\begin{equation}
 Y_* = -{D\over{2(D+2)}}~\lambda^{-1}~,
\end{equation}
which relates the mass parameter $m$ to the higher derivative coupling $\lambda$:
\begin{equation}
 m^2 = Y_* / D = -{1\over{2(D+2)}}~ \lambda^{-1}~.
\end{equation}
The graviton mass in this case is given by:
\begin{equation}
 M_h^2 = -{2\over{(D+2)^2}}~\lambda^{-1}~.
\end{equation}
Note that we must have $\lambda < 0$. Moreover, the cosmological constant $\Lambda$ must be tuned as follows:
\begin{equation}\label{tune.12}
 \Lambda = {{D^2 + 4D - 8}\over {4(D+2)^2}}~\lambda^{-1}~,
\end{equation}
which implies that the cosmological constant must be negative.

\subsection{Constrained Gravity as the Massless Limit}

{}Here we would like to begin addressing the question of what happens in the limit where the graviton mass goes to zero.
A naive expectation that one recovers Einstein-Hilbert gravity in this limit does not hold. Instead, we get {\em constrained} gravity.

{}To see what happens in the massless limit, for illustrative purposes, let us consider the example with the quartic
potential given by (\ref{quartic}). (The analysis in the general case completely parallels our
discussion here, so we will not repeat it.)
To obtain the massless graviton limit, we must take $\lambda\rightarrow -\infty$. In this limit both $Y_*$ and
$\Lambda$ go to zero, and so does $m^2$. Let us gauge fix the scalars to zero via (\ref{solphiY}). We then have:
\begin{equation}
 Y_{MN} = m^2 E_{MN}~.
\end{equation}
Here $E_{MN}$ is the Minkowski metric in the coordinate frame given by $x^M$. It coincides with the flat Minkowski metric $\eta_{MN}$
if $x^M$ are Minkowski coordinates. However, in the general case the metric $E_{MN}$ is not flat. For instance, in spherical coordinates
we have
\begin{equation}
 E_{MN}dx^M dx^N = -dt^2 + dr^2 + r^2\gamma_{ab} dx^a dx^b~,
\end{equation}
where $\gamma_{ab}$ is a metric on the unit sphere $S^{d-1}$, $d\equiv D-1$.

{}In the massless limit we then have the following equations of motion:
\begin{eqnarray}
 \label{const.1}
 &&\partial_M\left[\sqrt{-G} Q G^{MN}E_{NS}\right] - {1\over 2}\sqrt{-G} Q G^{MN}\partial_S E_{MN} = 0~,\\
 \label{einstein.1}
 &&R_{MN} - {1\over 2}G_{MN} R = 0~,
\end{eqnarray}
where
\begin{equation}\label{Q}
 Q\equiv 1 - {1\over{D+2}} ~G^{MN}E_{MN}~.
\end{equation}
Note that (\ref{const.1}) and (\ref{einstein.1}) are not linearized but exact. Eqn. (\ref{einstein.1}) is just the Einstein equation of
motion without the cosmological constant. However, gravity in the massless limit is {\em not} Einstein-Hilbert gravity as we have a
constraint (\ref{const.1}).

{}The fact that we obtain {\em constrained} gravity in the massless limit is important. In particular, if we take, say, a spherically
symmetric solution in massive gravity and consider the massless limit, we should {\em not} expect it to coincide with the Schwarzschild
solution of Einstein-Hilbert gravity. That would be like comparing apples to oranges. Instead, it should coincide with the
corresponding spherically symmetric solution in constrained gravity. The latter will look very different from the Schwarzschild
solution of Einstein-Hilbert gravity. One way to construct solutions in constrained gravity is to start with known solutions in
Einstein-Hilbert gravity and coordinate-transform them to satisfy the constraint (this is similar to \cite{GS}).
In other words, to compare apples to apples, we can, for example, compare the massless limit of a spherically
symmetric solution in massive gravity to an appropriately coordinate-transformed Schwarzschild
solution of Einstein-Hilbert gravity such that it satisfies the constraint.

\section{Spherically Symmetric Solutions}

{}Before we discuss explicit solutions, let us cover some generalities. We will be interested in spherically symmetric
solutions to massive and massless equations of motion ($A, B, C$ are functions of $r$ only):
\begin{equation}
 ds^2 = -A^2 dt^2 + B^2 dr^2 + C^2 \gamma_{ab} dx^a dx^b~.
\end{equation}
In the massive case, the Einstein equations read (for definiteness, we focus on the quartic potential (\ref{quartic})):
\begin{equation}
 R_{MN} = m^2 Q E_{MN} +{1\over{D-2}}\left[\Lambda -\lambda Y^2 \right]G_{MN}~,
\end{equation}
where $Q$ was defined in (\ref{Q}), $Y$ is given by
\begin{eqnarray}
 Y = m^2\left[ A^{-2} + B^{-2} + (D-2) r^2 C^{-2}\right]~,
\end{eqnarray}
and we gave $\Lambda$, $\lambda$ and $m^2$ in the previous section.

{}The non-vanishing components of $R_{MN}$
are given by (prime denotes derivative w.r.t. $r$):
\begin{eqnarray}
 &&R_{00} = A^2 B^{-2} \left[{A^{\prime\prime}\over A} - {A^\prime B^\prime\over {AB}} + (D-2) {A^\prime C^\prime\over {AC}} \right]~,\\
 &&R_{rr} = -\left\{{A^{\prime\prime}\over A} - {A^\prime B^\prime\over {AB}} + (D-2)
 \left[{C^{\prime\prime}\over C} - {B^\prime C^\prime\over {BC}}\right] \right\}~,\\
 &&R_{ab} = -\gamma_{ab} R_*~,\\
 &&R_*\equiv C^2 B^{-2} \left\{{C^{\prime\prime}\over C} + (D-3) \left({C^\prime\over C}\right)^2 +
 {C^\prime\over C}\left[{A^\prime\over A} - {B^\prime\over B}\right]\right\} - (D-3)~.
\end{eqnarray}
The scalar equations of motion (\ref{phiY.1}) reduce to a single equation:
\begin{equation}
 \partial_r\left[AB^{-1}C^{D-2} Q \right] - (D-2)r ABC^{D-4} Q = 0~.
\end{equation}
We will now discuss solutions to these equations. In $D=4$ we have some simplifications, so we will focus on $D=4$ for the remainder of this
paper.

\subsection{Four-dimensional Massless Solutions}

{}Let us start by analyzing the above equations in the massless case ($m^2 = 0$) in $D=4$. We have the following equations
\begin{eqnarray}
 &&{A^{\prime\prime}\over A} - {A^\prime B^\prime\over {AB}} + 2 {A^\prime C^\prime\over {AC}} = 0~,\\
 &&{A^{\prime\prime}\over A} - {A^\prime B^\prime\over {AB}} + 2
 \left[{C^{\prime\prime}\over C} - {B^\prime C^\prime\over {BC}}\right] = 0~,\\
 &&{C^{\prime\prime}\over C} + \left({C^\prime\over C}\right)^2 +
 {C^\prime\over C}\left[{A^\prime\over A} - {B^\prime\over B}\right] - B^2 C^{-2} = 0~,
\end{eqnarray}
plus the constraint
\begin{equation}\label{const-massless}
 \partial_r\left[AB^{-1}C^2 Q \right] - 2 r AB Q = 0~,
\end{equation}
where
\begin{equation}
 Q = 1 - {1\over 6}\left[A^{-2} + B^{-2} + 2r^2 C^{-2}\right]~.
\end{equation}
If it were not for the constraint, we could simply take the Schwarzschild solution:
\begin{eqnarray}
 &&{\overline A} = {\overline B}^{-1} = \sqrt{1-{r_*\over r}}~,\\
 &&{\overline C} = r~.
\end{eqnarray}
However, this solution does not satisfy the constraint.

{}Actually, there is a systematic way of fining solutions that satisfy the constraint by transforming known solutions that satisfy unconstrained
Einstein's equations. Thus, we start from a known unconstrained solution given by ${\overline A}, {\overline B}, {\overline C}$,
and transform the radial coordinate $r\rightarrow f(r)$. The resulting warp factors are given by
\begin{eqnarray}
 &&A(r) = {\overline A}(f(r))~,\\
 &&B(r) = {\overline B}(f(r)) f^\prime(r)~,\\
 &&C(r) = {\overline C}(f(r))~,
\end{eqnarray}
and they still satisfy the equations of motion. This is because the massless equations of motion possess full reparametrization invariance.
The constraint then produces a second order differential equation for the function $f(r)$. Thus,
starting with the Schwarzschild solution, we can obtain solutions satisfying the constraint by setting $f(r) = C(r)$ in the above
expressions, which gives a differential equation for $C$. We have:
\begin{eqnarray}
 A = \sqrt{1 - r_* / C}~,\\
 B = {C^\prime\over \sqrt{1 - r_* / C}}~,
\end{eqnarray}
and the differential equation for $C$ reads:
\begin{equation}\label{const-massless.11}
 \partial_r\left[A^2 C^2 Q / C^\prime \right] - 2 r Q C^\prime = 0~.
\end{equation}
While (\ref{const-massless.11}) is highly non-linear, we can solve it in two regimes: near the horizon ($C\rightarrow r_*$), and asymptotically ($r \gg r_*$).

\subsubsection{Asymptotic Behavior}

{}In the asymptotic regime we have $r \gg r_*$. We can find a solution via the following Ansatz:
\begin{eqnarray}
 C = r\left[1 + \alpha \left({r_*\over r}\right)^{1\over 2} + \beta~{r_* \over r} + {\cal O}\left({r_*\over r}\right)^{3\over 2}\right]~,
\end{eqnarray}
where $\alpha$ and $\beta$ are numerical coefficients to be determined. This indeed solves (\ref{const-massless.11}):
\begin{eqnarray}
 &&A = 1 - {r_*\over 2r} + {\cal O}\left({r_*\over r}\right)^{3\over 2}~,\\
 &&B = 1 + \sqrt{8\over 39} \left({r_*\over r}\right)^{1\over 2} + {r_* \over 2r} + {\cal O}\left({r_*\over r}\right)^{3\over 2}~,\\
 &&C = r\left[1 + \sqrt{8\over 39} \left({r_*\over r}\right)^{1\over 2} + \beta~{r_* \over r} + {\cal O}\left({r_*\over r}\right)^{3\over 2}\right]~,
\end{eqnarray}
and $\beta$ is an integration constant. This is because we started with the Schwarzschild solution and transformed it via $r \rightarrow C(r)$.
The constraint (\ref{const-massless.11}) is a second order differential equation for $C$, whose solution contains two integration constants. However, because
we drop subleading terms, the resulting equation actually is effectively only a first order equation for $C$, so we have one integration
constant (and the second integration constant controls the subleading terms). It simply parameterizes the Schwarzschild solution in the transformed coordinate frame.

{}Note that the above asymptotic solution is non-perturbative in the following sense. If we write
\begin{equation}
 C = r (1 + c)~,
\end{equation}
in the asymptotic regime $c$ is small, so we can try to solve (\ref{const-massless.11}) in the linearized approximation. However, one finds no solution when keeping only
linear terms, so one must also keep quadratic terms in $c$.

\subsubsection{Near-horizon Behavior}

{}In the near-horizon regime we have $C\rightarrow r_*$. In fact, in this case we have $r\rightarrow 0$, and:
\begin{eqnarray}
 &&C = r_* + \gamma r + {\cal O}(r/r_*)^2~,\\
 &&A = \sqrt{\gamma r/r_*}\left[1 + {\cal O}(r/r_*)\right]~,\\
 &&B = \sqrt{\gamma r_* / r} \left[1 + {\cal O}(r/r_*)\right]~.
\end{eqnarray}
These warp factors satisfy (\ref{const-massless.11}) for any positive integration constant $\gamma$. The near-horizon metric reads:
\begin{equation}
 ds^2 = -{\gamma r\over r_*} dt^2 + {\gamma r_*\over r} dr^2 + (r_* + \gamma r)^2 \gamma_{ab} dx^a dx^b~.
\end{equation}
This is the metric we ought to compare the massless limit of massive solutions to, and {\em not} to the Schwarzschild metric, in the
near-horizon regime.

\subsection{Four-dimensional Massive Solutions}

{}Let us now go back to the massive equations of motion. In four dimensions we have:
\begin{eqnarray}
 \label{EOM-mass-A}
 &&A^2 B^{-2} \left[{A^{\prime\prime}\over A} - {A^\prime B^\prime\over {AB}} + 2 {A^\prime C^\prime\over {AC}} \right] =
 m^2\left\{A^2\left[1 - {X^2\over 24}\right] - \left[1 - {X\over 6}\right]\right\}~,\\
 \label{EOM-mass-B}
 &&{A^{\prime\prime}\over A} - {A^\prime B^\prime\over {AB}} + 2
 \left[{C^{\prime\prime}\over C} - {B^\prime C^\prime\over {BC}}\right] = m^2\left\{B^2\left[1 - {X^2\over 24}\right] - \left[1 - {X\over 6}\right]\right\}~,\\
 &&C^2 B^{-2} \left\{{C^{\prime\prime}\over C} +  \left({C^\prime\over C}\right)^2 +
 {C^\prime\over C}\left[{A^\prime\over A} - {B^\prime\over B}\right]\right\} - 1 = \nonumber\\
 &&m^2\left\{C^2\left[1 - {X^2\over 24}\right] - r^2\left[1 - {X\over 6}\right]\right\}~,
 \label{EOM-mass-C}
\end{eqnarray}
plus the scalar equations of motion:
\begin{equation}\label{const-mass}
 \partial_r\left[AB^{-1}C^2 (6 - X)\right] - 2 r AB (6 - X) = 0~,
\end{equation}
where
\begin{eqnarray}
 X \equiv A^{-2} + B^{-2} + 2r^2 C^{-2}~.
\end{eqnarray}
Just as in the previous subsection, here we are interested in studying solutions to these equations in the asymptotic and near-horizon regimes. We will discuss
perturbative asymptotic solutions first.

\subsubsection{Perturbative Asymptotic Solutions}

{}Here our goal is to discuss the vDVZ discontinuity in perturbative asymptotic solutions. Let
\begin{eqnarray}
 &&A = 1 + a~,\\
 &&B = 1 + b~,\\
 &&C = r(1 + c)~.
\end{eqnarray}
Here we assume that $a, b, c$ go to zero asymptotically, and in the equations of motion we keep only linear
terms in $a, b, c$. As we will see in a moment, this approximation breaks down for small graviton mass, hence the vDVZ discontinuity.

{}The above four equations of motion in the linearized approximation read:
\begin{eqnarray}
 &&a^{\prime\prime} + {2\over r} a^\prime = {m^2\over 3} \left[3a + b + 2c\right]~,\\
 &&a^{\prime\prime} + 2c^{\prime\prime} + {4\over r} c^\prime - {2\over r} b^\prime = {m^2\over 3} \left[a + 3b + 2c\right]~,\\
 &&c^{\prime\prime} +  {4\over r} c^\prime + {1\over r} \left[a^\prime - b^\prime\right] - {2\over r^2}\left[b - c\right] = {m^2\over 3} \left[a + b + 4c\right]~,\\
 &&a^\prime + 2 c^\prime - {2\over r}\left[b - c\right] = 0~.
\end{eqnarray}
The solution is given by:
\begin{eqnarray}
 &&a^{\prime\prime} + {2\over r} a^\prime = \mu^2 a~,\\
 &&b = -{a^\prime \over{\mu^2 r}}~,\\
 &&c = -{1\over 2} (a + b)~,
\end{eqnarray}
where
\begin{eqnarray}
 \mu^2 = {2m^2\over 3}
\end{eqnarray}
is the mass squared of the graviton. Note that we have
\begin{equation}
 a + b + 2c = 0~,
\end{equation}
which is simply a manifestation of the fact that the trace of the graviton is not a physical degree of freedom.

{}Thus, we have:
\begin{eqnarray}
 &&a = {\zeta\over r} {\rm e}^{-\mu r}~,\\
 &&b = {\zeta\over \mu^2 r^3}\left[1 + \mu r\right]{\rm e}^{-\mu r}~,\\
 &&c = -{\zeta\over 2 \mu^2 r^3}\left[1 + \mu r + \mu^2 r^2\right]{\rm e}^{-\mu r}~,
\end{eqnarray}
where $\zeta$ is an integration constant. The only way to have a smooth massless limit would be to take $\mu\rightarrow 0$ and
$\zeta\rightarrow 0$ with $|\zeta|/\mu^2 \equiv r_1^3$ fixed. In this case we would have vanishing $a$ in the massless limit, {\em i.e.}, only $b$ and $c$ would be non-vanishing
\begin{eqnarray}
 &&a = 0~,\\
 &&b = -2c = {r_1^3\over r^3}~.
\end{eqnarray}
However, the corresponding
metric is equivalent to a coordinate-transformed flat metric (in spherical coordinates).

{}Above we assumed that $A$, $B$ and $C/r$ all asymptotically go to 1. In fact, there are no asymptotic solutions other than those where $A$, $B$ and $C/r$ all go
to 1. To see this, let us assume that $A$, $B$ and $C/r$ asymptotically go to some constant values ${\widetilde A}$, ${\widetilde B}$ and ${\widetilde C}$. Eqn.
(\ref{const-mass}) then implies that ${\widetilde B} = {\widetilde C}$, and the equations (\ref{EOM-mass-A}) and (\ref{EOM-mass-B}) imply
that ${\widetilde A} = {\widetilde B}$. Noting that asymptotically we then have $X \rightarrow 4 / {\widetilde A}^2$, Eqn. (\ref{EOM-mass-A}) implies that
${\widetilde A} = 1$.

{}Thus, as we see, {\em perturbative} 
asymptotic massive solutions do not have a smooth massless limit (except for trivial solutions equivalent to coordinate-transformed
flat solutions). This is the well-known vDVZ discontinuity \cite{vDV, Zak}. However, this discontinuity is an artifact of the perturbative approximation, which breaks
down at $r \sim r_1$. Note that $|\zeta|$ is expected to be of order of the Schwarzschild radius $r_*$, so we have $r_1\sim (r_* / \mu^2)^{1/3}$. This scale goes to infinity when $\mu$ goes to zero, so one must consider non-perturbative solutions \cite{Vain}.

\subsubsection{Non-perturbative Asymptotic Solutions}

{}The correct way to think about the massless limit is therefore to consider non-perturbative massive solutions. In the massless limit they smoothly go to the asymptotic massless solutions we discussed in Subsection 3.1.2. Thus, we have:
\begin{eqnarray}
 &&A = 1 - {r_*\over 2r} + {\cal O}\left({r_*\over r}\right)^{3\over 2} + {\cal O}(\mu^2\sqrt{r_* r^3})~,\\
 &&B = 1 + \sqrt{8\over 39} \left({r_*\over r}\right)^{1\over 2} + {r_* \over 2r} + {\cal O}\left({r_*\over r}\right)^{3\over 2} + {\cal O}(\mu^2\sqrt{r_* r^3})~,\\
 &&C = r\left[1 + \sqrt{8\over 39} \left({r_*\over r}\right)^{1\over 2} + \beta~{r_* \over r} + {\cal O}\left({r_*\over r}\right)^{3\over 2}
 + {\cal O}(\mu^2\sqrt{r_* r^3})\right]~,
\end{eqnarray}
where $\beta$ is an integration constant. Note that the expansion in $\mu^2$ is valid at distance scales $r\ll 1/\mu$. As $\mu\rightarrow 0$, we have a smooth massless limit for all $r$.

\subsubsection{Singular Solutions}

{}Let us finally comment on what happens in massive solutions in the analog of the near-horizon regime in the massless case. Recall that in the massless case at the horizon, which in the coordinate-transformed Schwarzschild solution is located at $r\rightarrow 0$, we had $A\rightarrow 0$, $B\rightarrow\infty$, $C\rightarrow r_*$,
with vanishing scalar curvature. In the massive case the situation is different. The scalar curvature is given by:
\begin{equation}
 R = m^2(X - 4)~.
\end{equation}
If $A$ goes to zero, then $X$ diverges, and so does $R$, which implies that if $A$ goes to zero, we have a naked singularity instead of a horizon.

{}Here we can try to look for solutions where $A$ goes to a constant instead. In this case $X$ goes to a constant as well, so the scalar curvature is
finite. However, here we argue that such solutions cannot reproduce the massless near-horizon behavior in the massless limit.

{}Thus, let us assume that in the massive case $A$ goes to a constant $A_0$ at the horizon. To have a smooth massless limit we must assume that $A_0$
goes to zero in the massless limit, and $X_0$, the value of $X$ at the horizon, then goes to infinity (which is the behavior we have in the massless case).
For sufficiently small graviton mass we can then assume that $X_0\neq 6$.

{}Next, (\ref{const-mass}) can be written as:
\begin{equation}
 {A^\prime\over A} - {B^\prime\over B} + {2C^\prime \over C} - {X^\prime\over{6-X}} = 2r~{B^2\over C^2}~.
\end{equation}
In the near-horizon regime we are assuming that $C$ also goes to a constant $C_0$. We then have the following leading behavior for $B$:
\begin{equation}\label{B.0}
 B \sim {C_0 \over 2\sqrt{r_0(r - r_0)}}~,
\end{equation}
where $r_0 > 0$ is the location of the horizon. Recall that in the massless case the horizon is located at $r = 0$.
If we assume that $r_0 = 0$, then we have $B$ divergent as $1/r$ instead of $1/\sqrt{r}$ (which is
the near-horizon behavior in the massless case). Moreover,
according to (\ref{B.0}) we cannot assume that $r_0$ goes to zero in the massless limit. Therefore, solutions where $A$ goes to a constant
at the horizon cannot reproduce the massless solutions in the massless limit.

\section{Non-singular Massive Solutions}

{}As we saw in the previous section, there is a smooth massless limit in the asymptotic regime in the massive spherically symmetric solutions. On the other hand,
massive spherically symmetric solutions are singular\footnote{This is consistent with the numerical results of \cite{Damour}, albeit the class of models studied there
was not obtained via gravitational Higgs mechanism, {\em i.e.}, diffeomorphisms were broken explicitly.}.
This simply means that the approximation we are using breaks down.
In particular, appearance of a naked singularity implies that higher curvature terms become important and should be included.
Conceptually, this is similar to the appearance of naked singularities in higher codimension brane solutions in Einstein-Hilbert gravity \cite{Greg}.
Once we include higher curvature terms, we expect to find smooth solutions, in particular, by using the smoothing out procedure of \cite{CIKL}.
In the context of higher codimension branes explicit non-singular solutions were indeed found in \cite{CIK}.

{}Similarly, we expect that non-singular spherically symmetric solutions also exist in the context of massive gravity once we add higher curvature terms. The latter
generally make the problem technically challenging. However, there are special higher curvature combinations, in particular, (lower-dimensional) Euler invariants, one can add that are sufficient to get rid of singularities, yet are simple enough to keep the problem computationally tractable.

{}Thus, we can add a Gauss-Bonnet combination of quadratic in curvature terms, whose simplifying property is that all equations of motion still remain
second order differential equations and no third or forth derivatives appear. However, the Gauss-Bonnet combination is topological in $D = 4$, so for it to contribute we
must consider $D > 4$. In fact, in $D = 5$ completely smooth spherically symmetric analogues of the Schwarzschild solution were explicitly constructed in \cite{IK} for massless Einstein-Hilbert-Gauss-Bonnet gravity (EHGBG) in $D=5$ in a different context.
We therefore expect non-singular spherically symmetric solutions in massive (via gravitational Higgs mechanism) EHGBG in $D = 5$ to exist, and perhaps even be computationally tractable. Note that adding higher curvature terms only affects high curvature regions, so it does not modify the no vDVZ conclusion in the asymptotic regime, while it should allow for a smooth massless limit at all distance scales.

\subsection*{Acknowledgments}

I would like to thank Alberto Iglesias for discussions.

%%%\newpage

\end{document}